\documentclass[10 pt,a4paper]{article}

\usepackage[
  top=2.5cm,
  bottom=2.6cm,
  left=2.1cm,
  right=2.1cm
]{geometry}

\usepackage{xcolor}
\usepackage{amsmath}
\usepackage{amssymb,amsfonts,amsthm}
\usepackage{graphicx}
\usepackage{float}
\usepackage{subfig,eurosym}
\usepackage{verbatim}
\usepackage{mathrsfs}

\usepackage[linkcolor=blue,citecolor=blue,urlcolor=blue,colorlinks=true]{hyperref}

\def\pa{\partial}

\def\m{\mathcal}

\graphicspath{{./fig_crop/}}
\renewcommand{\(}{\left(}
\renewcommand{\)}{\right)}
\renewcommand{\[}{\left[}
\renewcommand{\]}{\right]}

\def\m{\mathcal} 

\begin{document}

\title{\Large{\bf Virtual work, thermodynamic structure of the spacetime, \\ and black hole criticality}}

\author{Dumitru Astefanesei$^{1}$\footnote{Email: {\tt dumitru.astefanesei@pucv.cl}}~,~ Gonz\'alo Casanova$^{1}$\footnote{Email: {\tt gundisalvuzcasanova@gmail.com}}~, and Ra\'ul Rojas$^{2}$\footnote{Email: {\tt raulox.2012@gmail.com}} \\
\\\textit{$^{1}$Pontificia Universidad Cat\'olica de
Valpara\'\i so, Instituto de F\'\i sica,} \\\textit{Av. Brasil 2950, Valpara\'{\i}so, Chile}\\
\\\textit{$^{5}$Departamento de F\'{\i}sica, Universidad de Atacama,} \\\textit{Avenida Copayapu 485, Copiap\'o, Chile}}

\maketitle

\begin{abstract}
We propose a new way to relate the black hole thermodynamics and geometry by generalizing 
the Euclidean formalism to include `virtual geometries', which do not necessarily satisfy Einstein
equations. This provides a physically well motivated route to study black hole criticality and obtain 
the Landau–Ginzburg potential. We compute the `virtual thermodynamic potential' and show that 
it satisfies a modified quantum statistical relation that is compatible with the first law of black 
hole thermodynamics 
supplemented with an extra term, interpreted as virtual work in previous literature. The novelty is 
that, within our formalism, we can explicitly compute this term as the first derivative of the virtual 
thermodynamic potential with respect to the horizon radius that is considered as the order parameter. 
Imposing the physical condition that the first derivative vanishes is at the basis of the matching 
between the first law of black hole thermodynamics and (one of the) Einstein equations evaluated 
at the horizon. Interestingly, imposing the physical conditions that the second 
and third derivatives vanish, we can concretely study the criticality and existence of swallow tails. As 
a specific example, we apply this formalism to an exact $4-$dimensional asymptotically flat hairy black hole, namely 
the generalized Kaluza-Klein (KK) black hole when the dilaton potential is included, and show that 
it is thermodynamically stable and has a non-trivial critical behaviour corresponding to an inverted swallowtail.
\end{abstract}

\hypersetup{linkcolor=blue}

\newpage
\tableofcontents

\section{Introduction}
Explaining the thermodynamic nature of black holes must be an essential feature of 
any complete quantum theory of gravity. Providing tools similar with the ones used 
for non-gravitational quantum systems to investigate the thermodynamics, phase diagram, and criticality 
of black holes is essential for this line of inquiry. The seminal work of Gibbons and Hawking 
 \cite{Gibbons:1976ue} is an example of how 
the gravitational path integral and partition function  can be used to extract
thermodynamic and statistical insights about quantum gravity. The thermal 
aspects of black holes can then be studied by the Euclidean path integral that has two parts: 
the gravitational contribution, which is the Euclidean action evaluated  
on the spacetime geometry and quantum contribution, which is obtained by computing the partition 
function of the quantum fields living on this geometry.\footnote{Some issues related to 
the Euclidean Einstein-Hilbert action  are discussed in the recent paper, \cite{Horowitz:2025zpx}.}
In this context, the subtle relationship between thermodynamics and geometry is 
revealed: the temperature is the periodicity of the Euclidean 
time once the conical singularity is 
removed and thermodynamic potential is related to the regularized Euclidean action.  In this way, 
the quantum statistical relation in (grand) canonical ensemble \cite{Gibbons:1976ue}, which is compatible with the 
first law of thermodynamics when the black hole entropy is determined by $1/4$ of the event 
horizon area (in theories without higher derivative corrections) \cite{Bekenstein:1973ur,Bardeen:1973gs,Hawking:1975vcx}, can be obtained.\footnote{In the last years, there 
is also important progress in understanding the entropy of Hawking radiation, see \cite{Almheiri:2020cfm} for a nice review.}

The black hole thermodynamics makes sense only in a theoretical framework in which
the black holes are thermodynamically stable. This is not the case for black holes in flat spacetime. However, a scalar field with self-interaction thermodynamically stabilizes the black hole \cite{Astefanesei:2019mds, Astefanesei:2024wfj, Astefanesei:2019qsg, Astefanesei:2020xvn}. Intuitively, the dilaton potential plays the role of a box  and so, since the backreaction is relevant sufficiently close to the horion, only the small hairy 
black holes are thermodynamically stable. While this is a toy example, it could very well be that the dark matter has the same effect. Elucidating this aspect could be relevant for understanding the existence of supermassive black holes that can not be formed by gravitational collapse. One can imagine a scenario where, in the early universe, small black holes exist in a suitable environment that makes them thermodynamically stable, e.g. surrounded by dark matter. Then, after a long time, they can grow up to become supermassive if there is enough matter around they can absorb.

Various interesting attempts to understand the thermodynamic origin of gravity were put forward 
\cite{Jacobson:1995ab, Padmanabhan:2002sha,Padmanabhan:2009vy,Verlinde:2010hp}, though  a clear and 
complete understanding is still lacking. Particularly useful for our work is 
the proposal of Padmanabhan \cite{Padmanabhan:2002sha,Padmanabhan:2009vy} to interpret Einstein equation 
as a thermodynamic identity for a virtual displacement of the horizon. While this expression is compatible 
with the first law of thermodynamics up to a virtual work term, it is again completely local and it was derived 
quite differently from the usual first law of black hole thermodynamics. Since General Relativity is a classical 
field theory of gravity and the black hole thermodynamics is inherently a quantum effect, one would 
expect to understand the origin of the virtual term (and, also, the appearance of Planck constant) directly 
from the Euclidean formalism. This hints to the fact that the `virtual geometries', which encode the horizon fluctuations of a black hole, could play a role in 
elucidating the puzzle of the relation between thermodynamics and dynamics (encoded in
Einstein equations) of the gravitational field.

Our main goal here is to provide another concrete application of Euclidean gravity action 
to study black hole criticality and obtain the Landau-Ginzburg potential. Interestingly, using 
this method, we are able to also solve the puzzle mentioned above. That is, we have to consider 
an infinite family of virtual geometries, evaluate the corresponding regularized Euclidean action (including counterterms), interpret in a consistent way the virtual work term, and only then impose the physical constraint. 
 From all virtual geometries, only the one that satisfies Einstein equations evaluated at the horizon is 
compatible with the first law of thermodynamics when the virtual work vanishes. The relevance of the horizon becomes clear by the fact 
that its radius, which is used as the order parameter (that distinguishes between different phases) 
in our formalism, controls the virtual work term. 

Specifically, our proposal resembles the virtual work techniques used in solid bodies. A 
kinematically admissible displacement field is defined to be one that does not violate the 
boundary conditions. Such a displacement field would induce some stress 
field within the body, though it might not satisfy the equations of motion. The 
virtual displacements, even if physically impossible, must be compatible with the geometry 
of the original structure, which in our case is the spacetime geometry itself. We therefore 
consider changing the black hole metric by virtual transformations, which preserve the asymptotic  
structure and existence of the horizon, but not the black hole temperature. In this way, we 
provide a general formalism to study the black hole criticality. 

Finally, as a concrete application, we are going to apply this formalism to an exact hairy black hole solution, namely a generalization of 
the Kaluza-Klein black hole \cite{Gibbons:1985ac} for which the dilaton is endowed with a potential that originates from an electromagnetic Fayet-Iliopoulos (FI) term in $N = 2$ extended supergravity in four spacetime 
dimensions. This solution is part of a more general class of exact hairy black hole solutions \cite{Anabalon:2012ta}, which can 
be embedded in supergravity \cite{Anabalon:2017yhv,Anabalon:2020pez} and exist also in flat spacetime \cite{Anabalon:2013qua}. 
Surprisingly, these asymptotically flat hairy black hole solutions are thermodynamically {\cite{Astefanesei:2019mds,Astefanesei:2024wfj}
and dynamically \cite{Astefanesei:2019qsg,Astefanesei:2020xvn} stable. We prove that there 
exists a critical point in the grand canonical ensemble and find that 
the corresponding critical behaviour is of an inverted swallowtail with the stable black hole 
region appearing on the lower part of the curve instead of the upper part.

\section{Virtual thermodynamic potential}
\label{sec2}

\subsection{The model}

The theory under consideration is an Einstein-Maxwell-dilaton model endowed with a particular supergravity dilaton potential \cite{Anabalon:2017yhv,Anabalon:2020pez}. It 
is described by the action:
\begin{equation}
\label{action}
I_{\text{bulk}}=\frac{1}{16\pi}\int_{\mathcal{M}}{d^4x\sqrt{-g}\[R-e^{\gamma\phi} F^2 -\frac{1}{2}(\pa\phi)^2-V(\phi)\]}\equiv \frac{1}{16\pi}\int_{\mathcal{M}}d^4x\sqrt{-g}L
\end{equation}
where $\phi$ is the scalar field with self-interaction potential $V(\phi)$, and $(\pa\phi)^2\equiv g^{\mu\nu}\pa_\mu\phi\pa_\nu\phi$. The field strength is defined as $F_{\mu\nu}=\pa_\mu A_{\nu}-\pa_\nu A_{\mu}$, where $A_\mu$ is the gauge potential, and $F^2\equiv F_{\mu\nu}F^{\mu\nu}$. We have also adopted the natural units 
$G=c=1$, so that the gravitational coupling constant is $\kappa=8\pi$, and $\epsilon_0=(4\pi)^{-1}$. 

While, in principle, our proposal of integrating the Euclidean action can be used for this general 
theory, in what follows we shall be focusing on the concrete case $\gamma=\sqrt{3}$ when the dilaton is endowed with the potential discussed in \cite{Anabalon:2017yhv,Anabalon:2020pez}, which has the form:
\begin{equation}\label{pot}
V(\phi)=\alpha\[\sinh\Big(\sqrt{3}\phi\Big)
+9\sinh\(\frac{\phi}{\sqrt{3}}\)
-\frac{12 \phi}{\sqrt{3}}\cosh\(\frac{\phi}{\sqrt{3}}\)\]
\end{equation}
where $\alpha$ is a dimensionful parameter, with dimensions of $\ell^{-2}$, where $\ell$ is the dimensional 
unit of length.
It was shown that the model (\ref{action}) with the potential (\ref{pot})  is a consistent truncation of $N = 2$ supergravity in four
spacetime dimensions, coupled to a vector multiplet and deformed by a Fayet-Iliopoulos
term \cite{Anabalon:2017yhv, Anabalon:2020pez}.

The advantage of considering the dilaton superpotential  (\ref{pot}) is that the theory supports an exact asymptotically flat hairy black hole solution \cite{Anabalon:2013qua} (the generalization of KK black hole) that, surprisingly, is thermodynamically stable \cite{Astefanesei:2019mds, Astefanesei:2024wfj, Astefanesei:2019qsg, Astefanesei:2020xvn}.

\subsection{Virtual geometry}
A deformation of the black hole horizon by a virtual transformation is defined as a `virtual geometry' 
that preserves the boundary conditions and existence of the horizon, though its temperature is arbitrary. The 
`virtual thermodynamic potential' is obtained from the regularized Euclidean action, including the counterterms 
\cite{Mann:1999pc,Lau:1999dp,Mann:2005yr},\footnote{We 
emphasize that, since  the asymptotic value of the dilaton is fixed by 
the potential, there is no need of extra counterterms, which should 
take care of such a variation \cite{Astefanesei:2018vga}.} evaluated on a 
virtual geometry and is compatible with a modified first law of black hole  thermodynamics that 
includes a virtual work term. 

Let us consider a general asymptotically flat spacetime geometry, which corresponds to the following static spherically symmetric ansatz\footnote{The ansatz (\ref{ansatz}) has been extensively considered in the context of hairy black holes,
e.g. \cite{Anabalon:2012ta, Anabalon:2017yhv, Anabalon:2020pez, Anabalon:2013qua,Acena:2012mr,Acena:2013jya,Anabalon:2013eaa, Anabalon:2024qhf, Anabalon:2025sqr}.}
\begin{equation}
\label{ansatz}
ds^2=\Omega(x)\[- f(x)dt^2+\frac{\eta^2dx^2}{ f(x)}+d\theta^2+\sin^2\theta d\varphi^2\]
\end{equation}
where $x$ is a dimensionless coordinate related to the usual radial one, which is compatible with the canonical asymptotically flat Arnowitt-Deser-Misner (ADM) structure \cite{Arnowitt:1960es,Arnowitt:1960zzc,Arnowitt:1961zz,Arnowitt:1962hi}, by the following transformation:
\begin{equation}
x=1+\frac{1}{\eta r}+\cdots
\end{equation}
with $x>1$, where $x\to 1$ represents the location of the boundary at $r\to\infty$ and $x\to \infty$ 
corresponds to $r=0$ (e.g., the singularity of the hairy black hole). For computational reasons, we 
have also introduced in a non-usual way the integration constant $\eta$ with the dimensional 
unit $\dim{\eta}=\ell^{-1}$. As we will show in Section \ref{sec4}, while such a rescaling is trivial for Reissner–Nordstr\"om
black hole, it can be interpreted as one of the integration constants for charged hairy black holes.  The metric function $f(x)$ has then the dimensional unit $\dim{f}=\ell^{-2}$, such that the expression in the parentheses is dimensionless.  Without loss of generality, we assume $\eta > 0$.

The metric functions should consistently behave at the boundary as 
\begin{equation} \label{falloff}
\Omega(x)=r^2+\cdots=\frac{1}{\eta^2(x-1)^2}+\cdots\,, 
\qquad -g_{tt}=f(x)\Omega(x)= 1+\cdots 
 \;\Rightarrow\; f(x)=\eta^2(x-1)^2+\cdots
\end{equation}

Since the dilaton equation of motion, 
\begin{equation}
\label{dilaton}
\mathcal{E}_\phi\equiv\nabla_\mu\nabla^\mu\phi-\frac{dV(\phi)}{d\phi}-\sqrt{3}e^{\sqrt{3}\phi}F^2=0
\end{equation}
is not an independent equation for the metric (\ref{ansatz}), 
the full system
of equations is determined by the independent Einstein equations $\mathcal{E}^\mu_\nu=0$, where  
\begin{equation}\label{eins}
\mathcal{E}_{\mu\nu}\equiv R_{\mu\nu}-\frac{1}{2}Rg_{\mu\nu}-\frac{1}{2}\pa_\mu\phi\pa_\nu\phi+\frac{1}{2}g_{\mu\nu}\[\frac{1}{2}(\pa\phi)^2+V(\phi)\]-8\pi T^{\text{EM}}_{\mu\nu}
\end{equation}
where $T^{\text{EM}}_{\mu\nu}\equiv \frac{1}{4\pi}e^{\sqrt{3}\phi}(F_{\mu\rho}F_{\nu}{}^{\rho}-\frac{1}{4}g_{\mu\nu}F^2)$.
Therefore, we have a system of three independent equations, namely $\mathcal{E}_t^t=\mathcal{E}_x^x=\mathcal{E}_\theta^\theta=0$, for three variable 
functions, namely $f(x),\, \phi(x)$, and $\Omega(x)$. In what follows, however, we will impose only two of these equations,
which determine $\phi(x)$ and $\Omega(x)$, while $f(x)$ will remain
unspecified.

To be concrete, we can interpret this geometry as a black hole with an event horizon at $x_h$ when $f(x_h)=0$, though it can be any geometry that contains 
a null surface. The advantage 
of using this specific ansatz is that it provides us exact asymptotically flat hairy black hole 
solutions in a nice analytic form \cite{Anabalon:2013qua}.

Lacking a satisfactory quantum description of the gravitational interaction, 
quantum field theory in curved spacetimes is the adequate framework to 
describe the interaction of quantum fields with the space-time curvature. 
The thermal aspects of black holes can be studied on the Euclidean 
section \cite{Gibbons:1976ue} obtained by `rotating' real time to imaginary time by 
the usual Wick rotation, $t\to -i\tau$. Once the quantum fluctuations are 
traded for thermal fluctuations, the temperature can be computed as 
the inverse of the periodicity of $\tau$. What is important for us is that 
we can still compute the temperature associated 
with virtual black hole geometry (\ref{ansatz}), there is no need of using Einstein equations. 
That is, under the physical assumption that the Euclidean geometry is smooth, the 
conical singularity  can be removed by demanding a periodicity $\beta$ of the Euclidean 
time, and so
\begin{equation}
\label{TildeT}
\beta\equiv\frac{\hbar}{k_BT}=\frac{4\pi\eta}{\left| f'(x_h)\right|} \quad\Rightarrow \quad T=\frac{\hbar}{k_B} \frac{|f'(x_h)|}{4\pi\eta}
\end{equation}
where $f'(x)\equiv \frac{d}{dx}f(x)$, $\hbar$ is the reduced Planck constant, and $k_B$ is 
the Boltzmann constant.

\subsection{Euclidean action and thermodynamics of virtual geometry}
We are interested in Einstein-Maxwell-dilaton theory (\ref{action}), where the metric (\ref{ansatz}) is  considered as a virtual geometry. The virtual displacement of the horizon is encoded in $f(x)$. From a physical point of view, the thermodynamic mass associated to a virtual displacement is the sum of a heat energy, namely the temperature
multiplied by the change in entropy, and the work done during this virtual deformation of the horizon. From a mathematical point of view, the virtual work can be computed as a horizon contribution of the Euclidean action evaluated on the virtual geometry.  

The total Euclidean action, including the Gibbons-Hawking boundary term and gravitational 
counterterm, is
\begin{equation}
\label{Eaction}
I=I^{\text{E}}_{\text{bulk}}
+I^{\text{E}}_{\text{GH}} 
+I^{\text{E}}_{\text{ct}} =I^{\text{E}}_{\text{bulk}}-\frac{1}{8\pi}
\int_{\pa\mathcal{M}}{d^3x\sqrt{h^\text{E}}K} +\frac{1}{8\pi}\int_{\pa\mathcal{M}}{d^3x\sqrt{h^\text{E}}\sqrt{2\mathcal{R}}}
\end{equation}
where $K$ is the trace of the extrinsic curvature tensor $K_{\mu\nu}\equiv h^{\;\alpha}_\mu h^{\;\beta}_\nu \nabla_\alpha n_\beta$ with  the unit normal to the hypersurface $x=\text{constant}$, given by $n_\mu=-\frac{\delta_\mu ^x}{\sqrt{g^{xx}}}$, and $\mathcal{R}=\frac{2}{\Omega(x)}$ is the Ricci scalar on that hypersurface with the induced metric $h_{ab}$. 

The gravitational counterterm, $I^{\text{\,E}}_{\text{ct}}$,  was added to remove the divergent contributions 
of the Euclidean action for asymptotically flat spacetimes \cite{Mann:1999pc,Lau:1999dp}. The quasilocal formalism supplemented with counterterms in flat spacetime and the corresponding regularized quasilocal stress tensor was used in \cite{Astefanesei:2005ad} to obtain the
thermodynamics of the dipole black ring \cite{Emparan:2004wy}. However, this formalism was put on a firm ground by Mann
and Marolf in \cite{Mann:2005yr}, where they have shown that the addition of an appropriate covariant boundary term to the
gravitational action yields a well-defined variational principle for asymptotically flat spacetimes. This becomes
the standard tool that leads to a natural definition of conserved quantities at spatial infinity, e.g. \cite{Astefanesei:2006zd,Astefanesei:2009wi,Compere:2011db,Compere:2011ve,Astefanesei:2010bm,Mann:2006bd,Mann:2008ay}.

We consider the exact hairy black hole solution associated to the dilaton potential 
(\ref{pot}) for which the dilaton and conformal factor are \cite{Anabalon:2013qua}
\begin{equation}\label{phiomega0}
	\phi(x)=\sqrt{3}\ln(x)\,, \qquad \Omega(x)=\frac{4x}{\eta^2(x^2-1)^2}
\end{equation}
The Maxwell equation can be integrated in the electrostatic regime yielding to
\begin{equation}
\label{elemgfield}
F=dA=q\exp\(-\sqrt{3}\phi\)dt\wedge dx
\end{equation}
Using (\ref{elemgfield}), the physical electric charge can be then expressed in terms 
of the integration constants  as,
\begin{equation}\label{charg}
Q(\eta, q)=\frac{1}{4\pi}\oint_{s_\infty^2}{\star\(e^{\sqrt{3}\phi} F\)}=\frac{q}{\eta}
\end{equation}
and further, using (\ref{phiomega0}), we can explicitly obtain the electric field as $F_{tx}=\eta Q/x^3$. We  fix the gauge such that  the electric potential vanishes at the horizon and matches 
the conjugate potential at the boundary:
\begin{equation}
A=\[\frac{q(1-x^2)}{2x^2}+\Phi\]dt \,, \qquad \Phi
=\frac{\eta Q(x_h^2-1)}{2x_h^2}
\end{equation}

Now, we evaluate the Euclidean action for an arbitrary function $f(x)$ corresponding to a virtual fluctuation of the horizon. The temperature (\ref{TildeT}) was computed, as usual,  by eliminating the conical singularity, and so the equations of motion are not used in any way in this computation. The conformal factor, obviously, do not enter in the expression of the temperature and so it can be fixed while the temperature can be still kept arbitrary. 

We define the virtual thermodynamic potential as
\begin{equation}\label{defG}
\mathcal{G}(T, \Phi, x_h)\equiv\beta^{-1}I^\text{E} 
=M-TS-\Phi Q
\end{equation}
where, within this formalism, we identify
\begin{equation}
\label{Vmass}
M(\eta,q)=\frac{1}{4\eta}\(\frac{x_h^2+3}{x_h^2-1} 
+\frac{x_h^2-1}{x_h^2}q^2\)
+\[\frac{x_h^4-8x_h^2-17}{6(x_h^2-1)^2} +\frac{2\ln{x_h}(3x_h^2+1)}{(x_h^2-1)^3}\]\frac{\alpha}{\eta^3}\end{equation}
as the `virtual thermodynamic mass' of the system, 
and $S\equiv\frac{k_B}{4\hbar}\mathcal{A}$, where $\mathcal{A}=4\pi\Omega(x_h)$ is the horizon area. We emphasize that the virtual thermodynamic mass is not the quasilocal 
mass of the physical solution. However, we are going to prove in Section \ref{sec4} that, when the virtual work term vanishes, the virtual themodynamic mass matches the quasilocal mass.

\section{Thermodynamic structure of the spacetime and the relation with the previous literature}
\label{sec3}

In this section, we prove that the thermodynamic relation presented 
in \cite{Padmanabhan:2002sha,Padmanabhan:2009vy}, which
contains a virtual work term, is equivalent with the first law of thermodynamics for 
the virtual geometry (\ref{ansatz}). With this new interpretation, there is no problem 
related to the existence of a non-zero electromagnetic energy-momentum tensor 
in the right hand side of Einstein equations when 
the extra electrostatic contribution should be added to the mass \cite{Padmanabhan:2009vy}.

In Section $5.1$ of \cite{Padmanabhan:2009vy} entitled `An unexplained connection between horizon 
thermodynamics and gravitational dynamics', Padmanabhan proved that Einstein
equations can be reinterpreted as a thermodynamic identity for a virtual displacement
of the horizon by an infinitesimal amount. Even if the entire 
description is local, this identity is compatible with the usual 
first law of black hole thermodynamics up to a virtual work term, 
which allows obtaining the expressions for energy and entropy.  Besides the appearance 
of the virtual term, one problem with this 
interpretation is that the Planck constant 
appears only in using the Euclidean extension of the metric to fix 
the form of temperature. Otherwise, it would be quite strange to 
use the dynamics of classical gravity to obtain the black hole thermodynamics 
that is quantum in origin. Due to the equivalence principle, another 
obvious problem is that the concept of localized energy is challenging to define. 
An understanding 
of the first law of black hole thermodynamics requires a definition of the thermodynamic 
potential (and energy) that is coming from a computation of the action including the boundary terms and so a complete apprehension of
thermodynamics, at the semiclassical level, demands relating computations at the horizon and 
black hole boundary.  

In what follows, we relate the results from the previous section with Padmanabhan's proposal 
\cite{Padmanabhan:2002sha,Padmanabhan:2009vy}. That is, we explicitly show that the virtual geometries are compatible 
with the thermodynamic identity arising out of infinitesimal virtual
displacements of the horizon. In other words, we are going to demonstrate that, 
in fact, this thermodynamic relation corresponds to a modified first law 
$dM - T dS -\Phi dQ = \delta W_{\text{virtual}}$  when 
considering the virtual thermodynamic potential. The usual first law of thermodynamics,
when $\delta W_{\text{virtual}}=0$, is recovered only when the Einstein equations are 
satisfied at the horizon. 

From the definition of the virtual thermodynamic potential (\ref{defG}), we obtain
\begin{equation}
\label{first1}
d\mathcal{G}=
\(\frac{\pa\mathcal{G}}{\pa T}\)_{\Phi,x_h}dT +\(\frac{\pa\mathcal{G}}{\pa\Phi}\)_{T,x_h}d\Phi +\(\frac{\pa\mathcal{G}}{\pa x_h}\)_{\Phi,T}dx_h=-SdT-Qd\Phi+\(\frac{\pa\mathcal{G}}{\pa x_h}\)_{\Phi,T}dx_h
\end{equation}
compatible with a modified first law of black hole thermodynamics:
\begin{equation}
\label{firstlaw}
dM-TdS-\Phi dQ=\(\frac{\pa\mathcal{G}}{\pa x_h}\)_{\Phi,T}dx_h
\end{equation}

We emphasize that, since  we 
have only fixed the scalar field and conformal 
factor in the metric, leaving $f(x)$ arbitrary, one of the equations of motion, e.g. $\mathcal{E}_t^t=0$ of (\ref{eins}), 
still remains unsolved. The essential observation is that the partial derivative of the virtual potential with respect to the horizon radius that appears on the right-hand side of eq. (\ref{firstlaw}) is 
\begin{equation}
\(\frac{\pa\mathcal{G}}{\pa x_h}\)_{{\Phi,T}}
=-\frac{\pi k_B}{\hbar}\Omega'(x_h)(T-T_0) =\frac{1}{2}{\eta\Omega^2(x_h)}\mathcal{E}_t^t(x_h)
\end{equation}
where
\begin{equation}
T_0\equiv\frac{\hbar}{2\pi k_B}\frac{\eta\Omega(x_h)}{\Omega'(x_h)}
\[\frac{q^2e^{-\sqrt{3}\phi_h}}{\eta^2\Omega(x_h)}
+\frac{\Omega(x_h)V(\phi_h)}{2}-1\]
\end{equation}
with $\phi_h\equiv \phi(x_h)$. Therefore, by imposing the on-shell condition 
at the horizon, $\mathcal{E}_t^t(x_h)=0$, 
we obtain the usual first law of black hole thermodynamics. 

While the mathematical expression of the first law for hairy black holes \cite{Gibbons:1996af}
is correct, its physical interpretation 
is problematic. This point was clarified in \cite{Astefanesei:2018vga}.

In string theory, by changing the asymptotic values of the moduli, the theory is changed and so 
the correct interpretation is that the same black hole configuration is considered when the couplings 
of the theory are changed. The black hole energy should change accordingly. On the other hand, 
from a General Relativity point of view, we indeed obtain a modified first law, though this result 
is still physically incorrect. There is no conserved charge associated with the hair and so the 
scalar charges should not appear in the first law. Once the correct variational 
principle is considered, it was shown in \cite{Astefanesei:2018vga}  that the quasilocal mass satisfies 
the usual first law without extra terms.

Importantly, we would like to point out that, since in the presence 
of the dilaton potential its asymptotic value, $\phi_{\infty}$, is fixed, the results of \cite{Gibbons:1996af} and \cite{Astefanesei:2018vga} do not apply to our case. However, 
if we consider the parameter  of the potential, $\alpha$, as a modulus and its variation,
we obtain a modified Smarr formula similar with the extended thermodynamics 
in AdS \cite{Astefanesei:2019ehu,Astefanesei:2023sep}  when the cosmological constant is variable:
\begin{equation}
M-2TS-\Phi Q+2\alpha B=\frac{2\Omega(x_h)}{\Omega'(x_h)}\(\frac{\pa\mathcal{G}}{{\pa x_h}}\)_{\Phi,T}
\end{equation}
where
\begin{equation}
B=\frac{(x_h^2-1)(17+8x_h^2-x_h^4)-12\ln{x_h}(3x_h^2+1)}{6\eta^3(x_h^2-1)^3}
\end{equation}
is the conjugate\footnote{
$B\equiv 
\left(\frac{\pa{M}}{\pa{\alpha}}\right)_{S,Q}$
is required because of the presence of a dimensionful parameter, $\alpha$, in the action \cite{Hu:2018njr}.} to the parameter $\alpha$  that control the self-interaction 
of the scalar field. This result should 
be again interpreted as the change in the energy of a specific black hole configuration 
under the moduli variation. On-shell, we recover the usual Smarr 
relation of hairy black holes with a dilaton potential, $M-2T_{\text{BH}}S-\Phi Q+2\alpha B=0$.

\section{Criticality of KK BH with a supergravity potential}
\label{sec4}

In what follows we are going to use the virtual thermodynamic potential formalism to study the 
criticality of KK black hole with a non-trivial dilaton potential. For simplicity, in this section we 
fix $\hbar=1$, $k_B=1$.

We consider the exact hairy black hole solution,
\begin{equation}
\label{function_f0}
f(x)=\alpha\[2\ln{x}+\frac{1}{2}(x^2-1)(x^2-3)\]+\frac{\eta^2}{4}(x^2-1)^2\[1-\frac{q^2}{x^2}(x^2-1)\]
\end{equation}
presented in \cite{Anabalon:2013qua} when $\gamma=\sqrt{3}$ (in Appendix, we provide a different way to obtain this exact solution by using the integrability of the action). We can use (\ref{function_f0}) to obtain the black hole temperature
\begin{equation}
\label{Tbh}
T_{\text{BH}}=\frac{(x_h^2-1)^2}{4\pi\eta x_h}\[\frac{\eta^2q^2(2x_h^2+1)}{4x_h^2}
-\frac{1}{2}\frac{\eta^2x_h^2}{x_h^2-1}-\alpha\]
\end{equation}

Interestingly, we can compute this temperature in a different way without using the full bulk solution. 
If we consider the left hand side of the dilaton equation (\ref{dilaton}) at the horizon
%
\begin{equation}
\mathcal{E}_\phi(x_h) \equiv\frac{4\pi\phi'(x_h)}{\eta\Omega(x_h)}\(T-T_\phi\)
\end{equation}
where
\begin{equation}
T_\phi\equiv -\frac{1}{4\pi}\frac{\eta\Omega(x_h)}{\phi'(x_h)}\[\left.\frac{dV(\phi)}{d\phi}\right|_{x_h}
-\frac{2\sqrt{3}q^2e^{-\sqrt{3}\phi(x_h))}}{\eta^2\Omega^2(x_h)}\]
\end{equation}
then, the condition $T_0 = T_\phi$ is equivalent with $f(x_h)=0$ and thus we can recover (\ref{Tbh}). 
In this case, the following relation is satisfied:
\begin{equation}
\mathcal{E}_\phi(x_h)
=\frac{2\Omega(x_h)\phi'(x_h)}{\Omega'(x_h)}\mathcal{E}_t^t(x_h)
\end{equation}
Moreover, solving for $\eta$ from $T_0 = T_\phi$ and replacing it into the virtual thermodynamic mass (\ref{Vmass}), we recover the correct quasilocal mass associated to the physical solution,
\begin{equation} \label{ADM mass}
M_{\text{quasilocal}} = \frac{q^2}{\eta} - \frac{1}{2 \eta} - \frac{4 \alpha}{3 \eta^3}
\end{equation}

In the previous section, we have obtained the virtual thermodynamic potential, $\mathcal{G}(T, \Phi, x_h)$. We have also shown that setting $(\pa\mathcal{G}/\pa{x_h})_{T,\Phi}=0$ places the system on-shell. We now assert that imposing the vanishing of the first three derivatives, i.e.,
\begin{equation}
\label{conditions0}
\(\frac{\pa\mathcal{G}}{\pa x_h}\)_{\Phi,\beta}=0\,, \qquad 
\(\frac{\pa^2\mathcal{G}}{\pa x_h^2}\)_{\Phi,\beta}=0\,, \qquad
\(\frac{\pa^3\mathcal{G}}{\pa x_h^3}\)_{\Phi,\beta}=0
\end{equation}
provides a criterion to obtain critical points, equivalent to the standard condition for criticality, $\pa{T}/\pa{x_h}=0$ and $\pa^2{T}/\pa{x_h^2}=0$. Solving the conditions (\ref{conditions0}), we find that the charged hairy black hole in the asymptotically flat spacetime under consideration has a critical point in the phase space of the grand canonical ensemble \footnote{The other critical thermodynamic quantities are well-defined: 
\begin{equation}\notag
M_c={0.5927}\alpha^{-\frac{1}{2}}\,, \quad S_c={0.00351}{\alpha}^{-1} \,, \quad Q_c={1.0046}\alpha^{-\frac{1}{2}}
\end{equation}
}
\begin{equation}
\label{conditions1}
T_c\approx 0.0617{\sqrt\alpha} \,, \quad \Phi_c\approx 0.70348\,, \quad \mathcal{G}_c\approx-\frac{0.1142}{\sqrt{\alpha}}
\end{equation}
Usually, the critical points are accompanied by normal swallowtails \cite{Landau:1937obd,Landau:1980mil,Ginzburg:1950sr} with a corresponding Landau-Ginzburg potential 
as schematically shown in Fig. \ref{p1}. However, in special situations, e.g. when there exist reentrant phase transitions  \cite{Altamirano:2013ane,Astefanesei:2021vcp} the swallowtail is inverted as it is shown in Fig. \ref{p2}. Since the swallowtail is flipped, the stable regions appear on the lower part of the curve instead of the upper part. This clearly indicates stable temperature regions for a system that would otherwise be unstable. 
\begin{figure}[t!]
\centering
\includegraphics[scale=0.38]{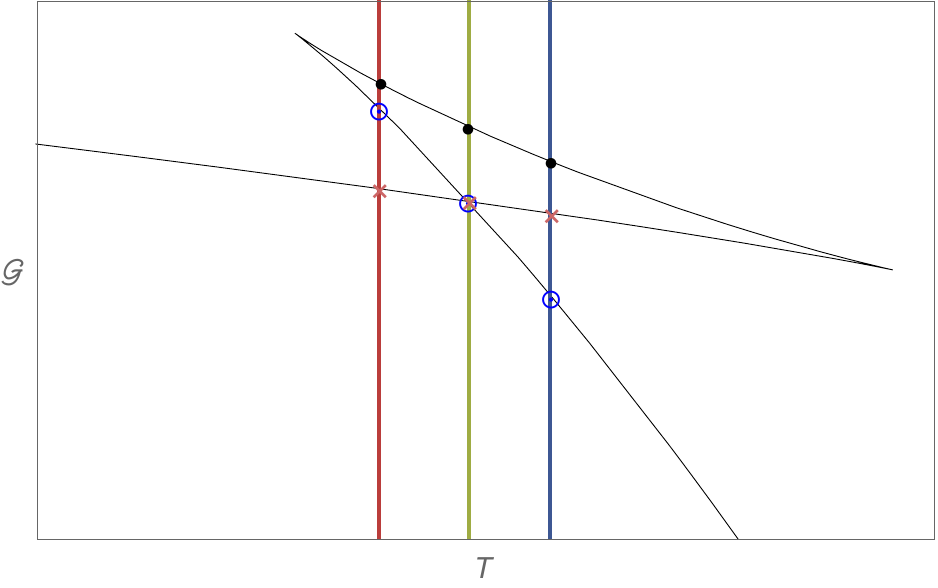}\quad
\includegraphics[scale=0.36]{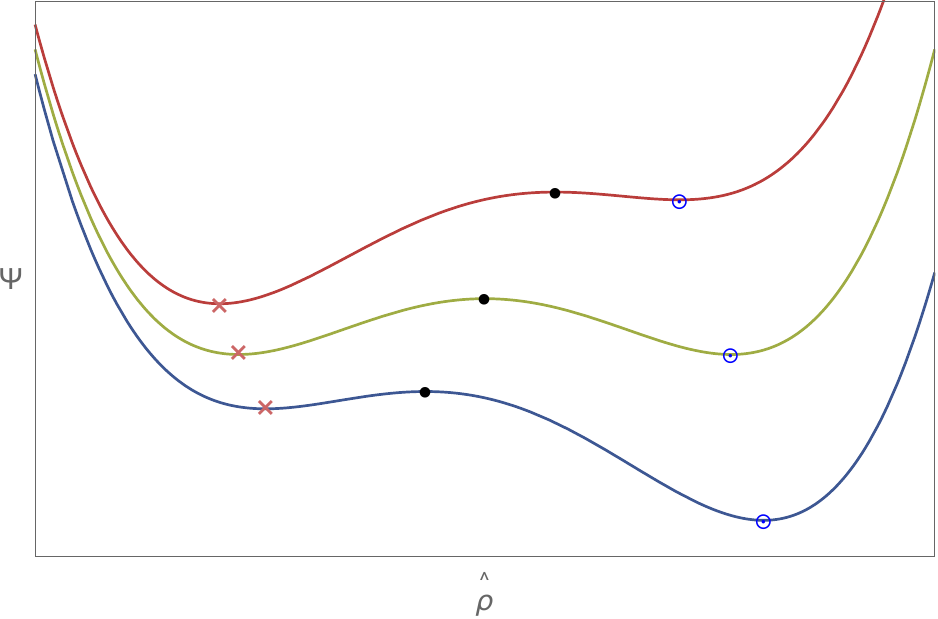}
\caption{\small \textbf{Top panel}: $\mathcal{G}$ vs $T$ displaying a standard swallowtail structure. \textbf{Bottom panel}: the corresponding Landau-Ginzburg potential, $\Psi$ vs $\hat\rho$, for three representative isotherms. For the leftmost isotherm (the red vertical line in the first plot), the globally stable phase corresponds to the branch of small black hole which corresponds to the global minimum of the uppermost curve in the bottom  panel. For the rightmost isotherm in top panel, the branch of large black hole phase is now the globally stable one.} 
\label{p1}
\end{figure}
\begin{figure}[t]
\centering
\includegraphics[scale=0.5]{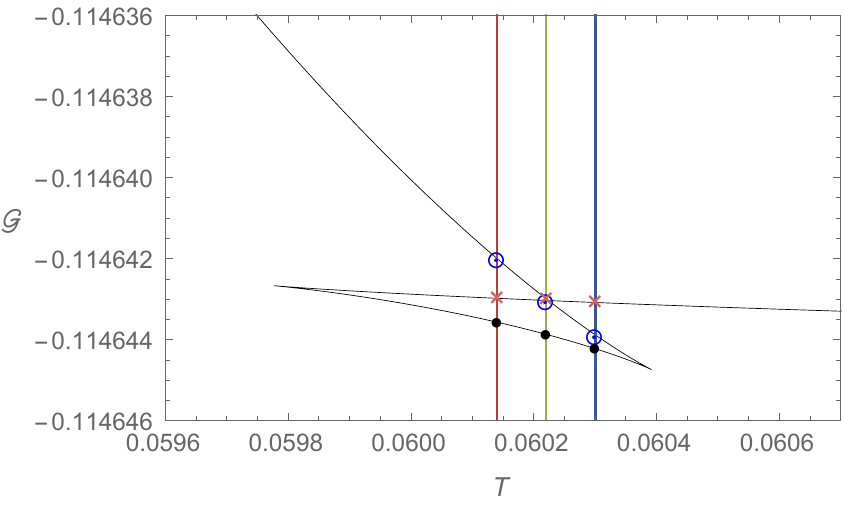}\quad
\includegraphics[scale=0.46]{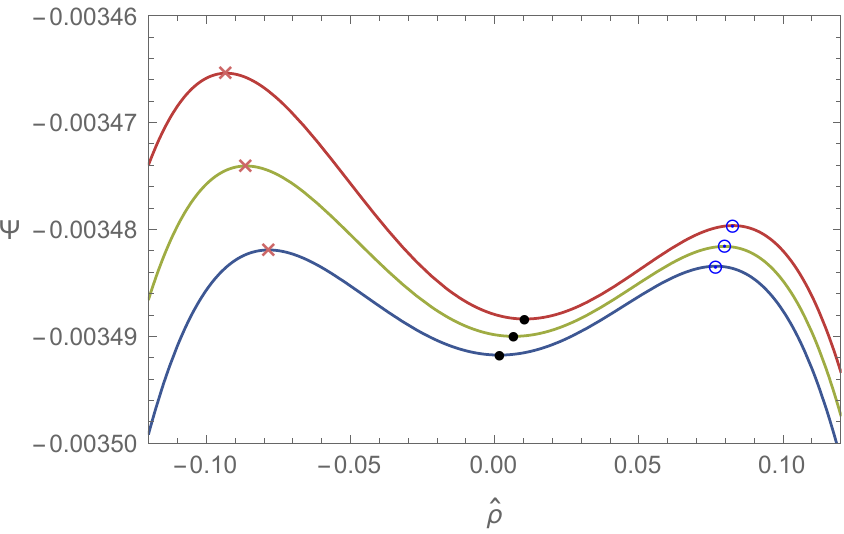}
\caption{\small \textbf{Top panel}: $\mathcal{G} = \mathcal{G}(T)$ for the value $\Phi = 0.70388$, for which the inverted swallowtail structure occurs. The cross, dot, and circle markers indicate small, intermediate, and large black holes, respectively, which intersect the isotherms $T \approx 0.0601$, $T \approx 0.0602$, and $T\approx 0.0603$.
\textbf{Bottom panel}: $\Psi = \Psi(\hat\rho)$ for $\hat{\psi} \approx 0.00057$ ($\Phi = 0.70388$), for the same three isotherms from the top panel.} 
	\label{p2}
\end{figure}
In Fig. \ref{p3}, we have depicted the behavior of the on-shell thermodynamic potential $\mathcal{G}(T,\Phi)$, where we can observe the appearance of inverted swallowtails. The dilaton potential is essential for the existence of thermodynamically stable black holes in flat spacetime. 
\begin{figure}[t!]
\centering
\includegraphics[scale=0.42]{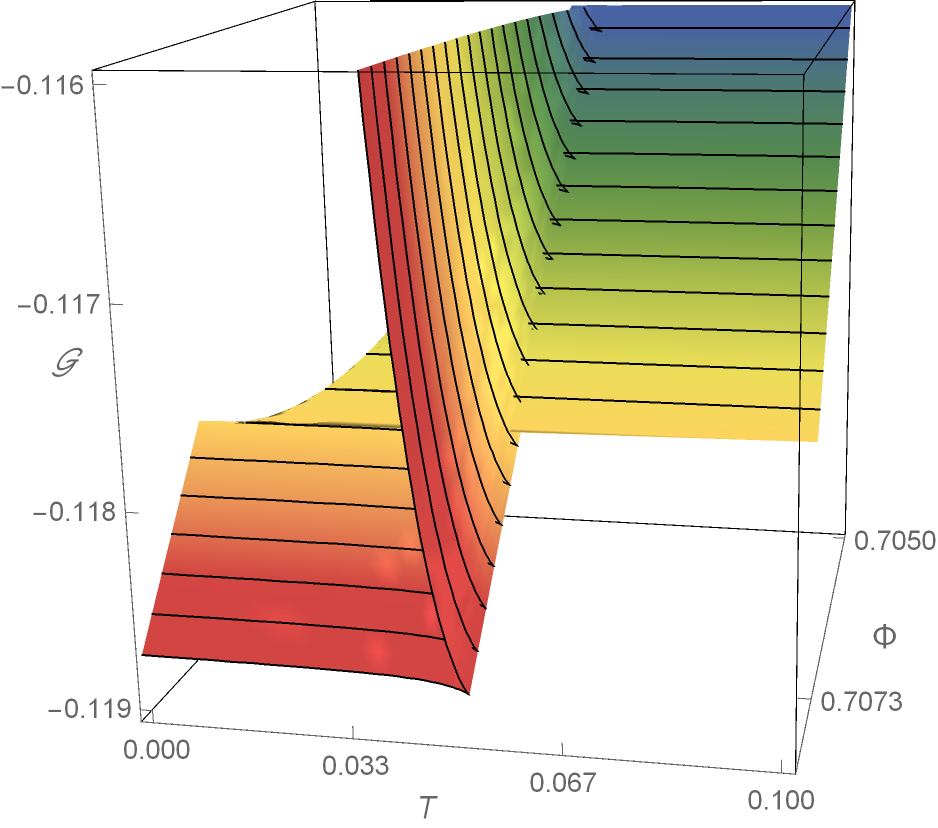}\quad
\includegraphics[scale=0.41]{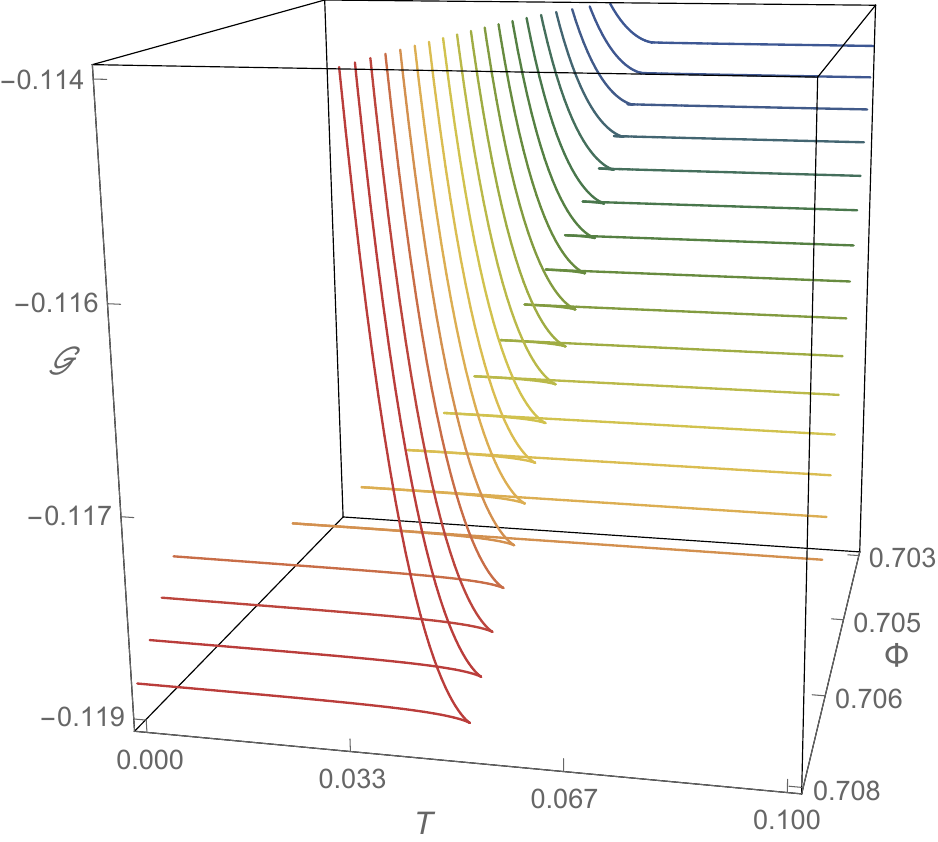}
\caption{\small \textbf{Left panel}: 3D plot of $\mathcal{G}$ vs $T$ and $\Phi$. \textbf{Right panel}: Constant-$\Phi$ curves. Near the critical point, the free energy surface develops a characteristic swallowtail structure, but in this case it appears inverted, with the `tail' pointing downward.}
\label{p3}
\end{figure}

To complete the analysis of the criticality, let us obtain the Landau-Ginzburg potential $\Psi$, by expanding the virtual thermodynamic potential around the critical point. 
The first step is to introduce the following local variables
\begin{equation}
\hat\rho\equiv  0.1643\(\frac{x_h}{x_c}-1\)-144.1874\hat\psi+4.7054\hat\beta\
\end{equation}
and
\begin{equation}
\hat\psi=\frac{\Phi}{\Phi_c}-1\,, \quad \hat\beta=\frac{\beta}{\beta_c}-1
\end{equation}
where $\beta=1/T$ is arbitrary. The critical point is now located at $\hat\rho=\hat\psi=\hat\beta=0$. In particular, the variable $\hat\rho$ plays the role of an order parameter, as shown below. The second step is to expand the virtual thermodynamic potential around the critical point up to the fourth order in $\hat\rho$,
\begin{equation}\label{landau}
\mathcal{G}\approx\mathcal{G}_c-0.029\hat\rho^4+(0.051\hat\beta-1.583\hat\psi)\hat\rho^2+(-0.004\hat\beta+0.172\hat\psi)\hat\rho+0.0002\hat\beta-0.707\hat\psi
\end{equation}
We now define $\Psi$, the Landau-Ginsburg potential, as
\begin{equation}\label{deflandau}
	\Psi \equiv \frac{\mathcal{G} - \mathcal{G}_c}{|\mathcal{G}_c|}
\end{equation}
%
In the expansion around the critical point, we obtain
\begin{equation}\label{landau3}
\Psi(\hat\rho)\approx -\frac{1}{4}\hat\rho^{\,4}
+\(0.4464\hat\beta-13.8596\hat\psi\)\hat\rho^2-\(0.0355\hat\beta-1.5033\hat\psi\)\hat\rho +0.0019\hat\beta-6.1857\hat\psi
\end{equation}
In Fig.\ref{p2}, we have depicted $\Psi$ vs $\hat\rho$, for a fixed $\hat\psi$ and for some values of $\hat\beta$.
The expression (\ref{landau3}) has the standard form of the Landau-Ginzburg potential presented in \cite{Goldenfeld:1992qy},
\begin{equation}
 \Psi(\hat\rho) = - \frac{\hat\rho^4}{4} - \tau \hat\rho^2 + H \hat\rho - C    
\end{equation}
where $H$ represents an external field, $\tau$ is the deviation of the critical temperature, and $\hat\rho$ is the order parameter. 

One important difference is the negative sign of the ${\hat\rho^4}$ term that is positive in the standard form. This suggests that there should be a higher-order term in the expansion with a positive coefficient to ensure that the potential is stable and bounded from below. However, it is important to concretely check first if that is indeed the case. Particularly, when there exist reentrant phase transitions the swallowtail is inverted and so it is useful to 
check the usual thermodynamic potential for the existence of stable black holes.

In Fig. \ref{p2}, three isotherms are depicted for a sensible value of $\Phi$. The lower part of the inverted swallowtail contains thermodynamically stable black holes. These configurations are located at the  
minimum of the Landau-Ginzburg potential for the corresponding values of $\hat\beta$. Unlike the AdS spacetime, there are no reentrant phase transitions for asymptotically flat hairy black holes, there exists only a phase of stable black holes that occurs within an interval $T_1(\Phi)\leq T\leq T_2(\Phi)$, for the range 
\begin{equation}
\Phi_c\approx 0.7035<\Phi<\frac{1}{\sqrt{2}}\approx 0.7071
\end{equation}
where the inverted swallowtail forms.

Since the Landau-Ginzburg potential was obtained using the virtual thermodynamic potential, the on-shell condition becomes (near the critical point) the equation of state,
 \begin{equation}\label{landau state}
    \(\frac{\pa \Psi}{\pa \hat\rho}\)_{\hat\beta,\hat\psi}\approx -\hat\rho^3-2 \tau\hat\rho+H= 0
\end{equation}
where the number of real roots for $\hat\rho$ is determined by the sign of discriminant $\Delta = -32\tau^3 -27H^2$.  In Fig. \ref{p2} we present a situation with $\Delta>0$, for which there are three real solutions, i.e., three configurations extremizing $\Psi$ (two maxima and one minimum), with the minimum representing the stable black hole configuration. In the cases $\Delta \leq 0$, there is a single real solution corresponding to one local maximum in $\Psi$ vs $\hat\rho$. In this situation swallowtails do not form and the configurations become all locally unstable.

\section{Conclusions}
We have provided a new formalism to study the black hole criticality. That is, we generalize the Euclidean formalism by including virtual geometries with 
an arbitrary temperature. Within this general framework, we explicitly complete the analysis of Padmanabhan 
\cite{Padmanabhan:2002sha,Padmanabhan:2009vy} and correctly interpret  the extra term in the first law of black hole thermodynamics as a virtual work term, which is related to the first derivative of 
the virtual thermodynamic potential with respect to the horizon radius. We obtain this 
result by evaluating the regularized Euclidean action, supplemented with boundary counterterms, on a virtual geometry corresponding to an arbitrary fluctuation of the horizon. Once we impose physical conditions on the virtual 
thermodynamic potential, we can obtain the critical points, investigate the existence of swallow tails 
by computing the associated temperatures, and determine the Landau-Ginzburg potential. 

While our proposal can be, in principle,  generally used for any gravity theory that contains black holes, 
a concrete computation of the virtual thermodynamic potential can be done only if some integrability 
conditions, which are specific to the chosen model, are imposed. This is why, for concreteness,  we 
have focused on a specific Einstein-Maxwell-dilaton theory with a dilaton potential obtained from 
the $N=4$ supergravity with FI terms \cite{Anabalon:2017yhv,Anabalon:2020pez}. The embedding in supergravity provides a consistent theoretical framework for which the theory has a well defined ground state. Interestingly, 
the theory contains analytic hairy black hole solutions that are thermodinamically stable and, unexpectedly, they have non-trivial criticality.
\begin{figure}[t!]
\centering
\includegraphics[scale=0.16]{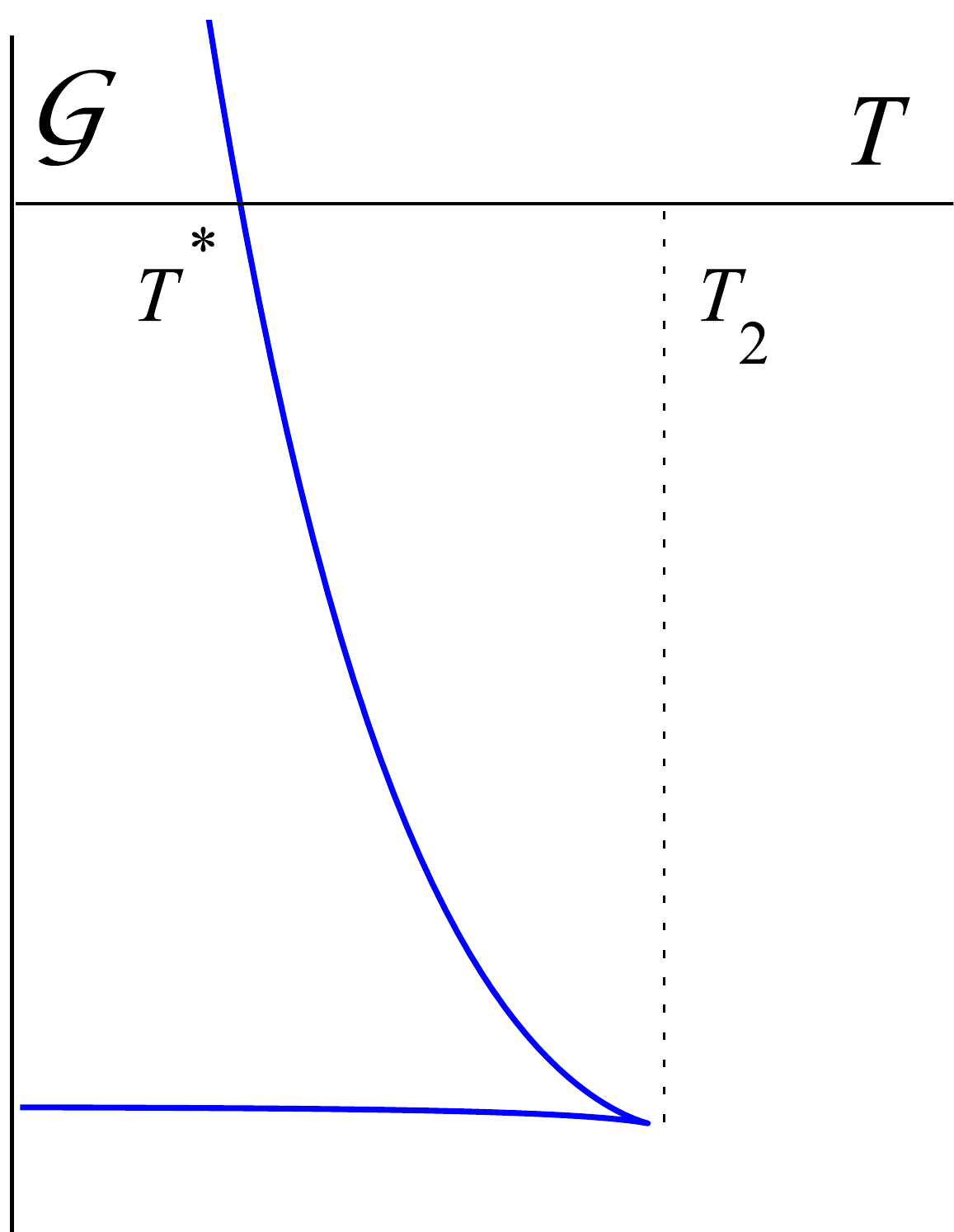}\qquad
\includegraphics[scale=0.16]{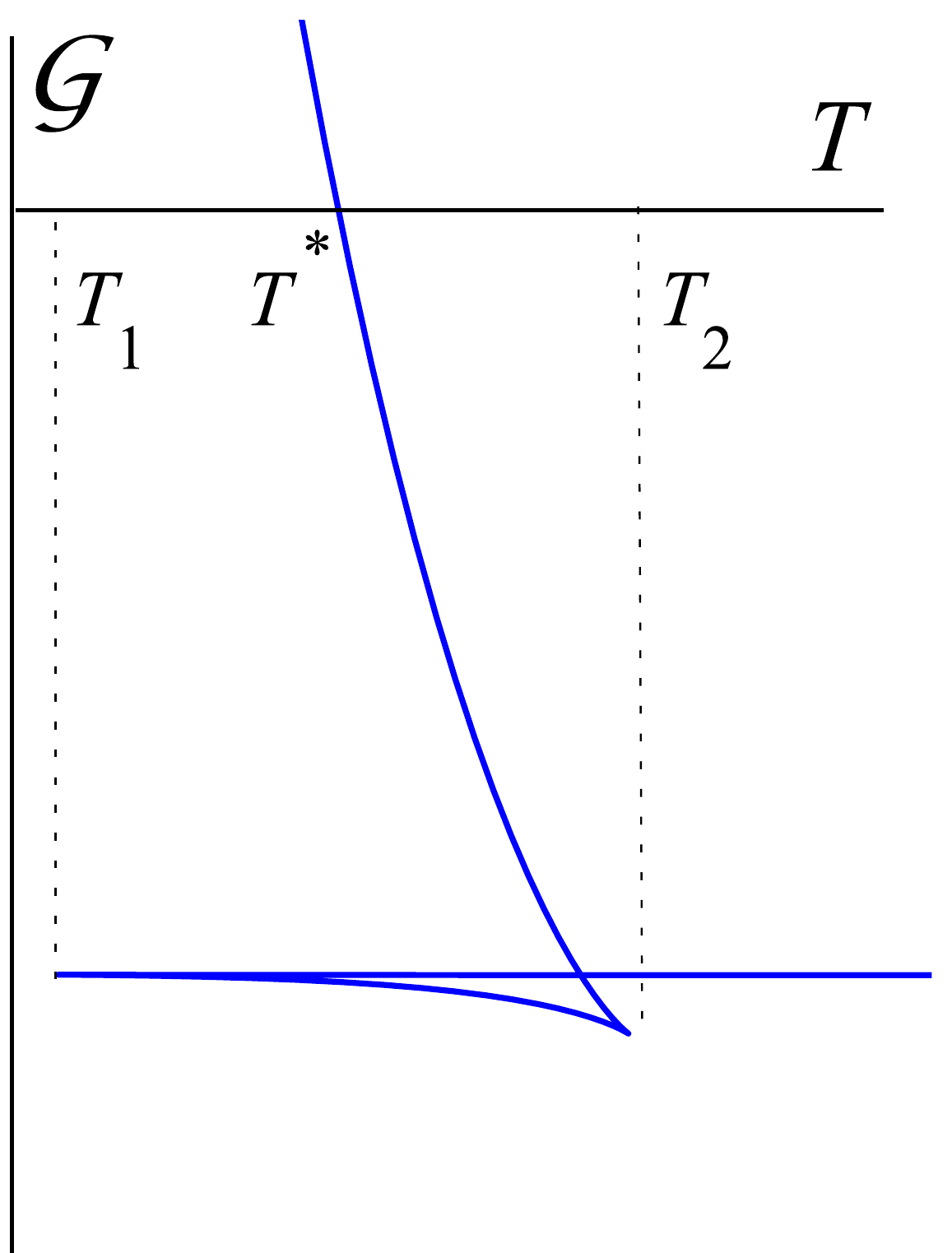}\qquad
\includegraphics[scale=0.16]{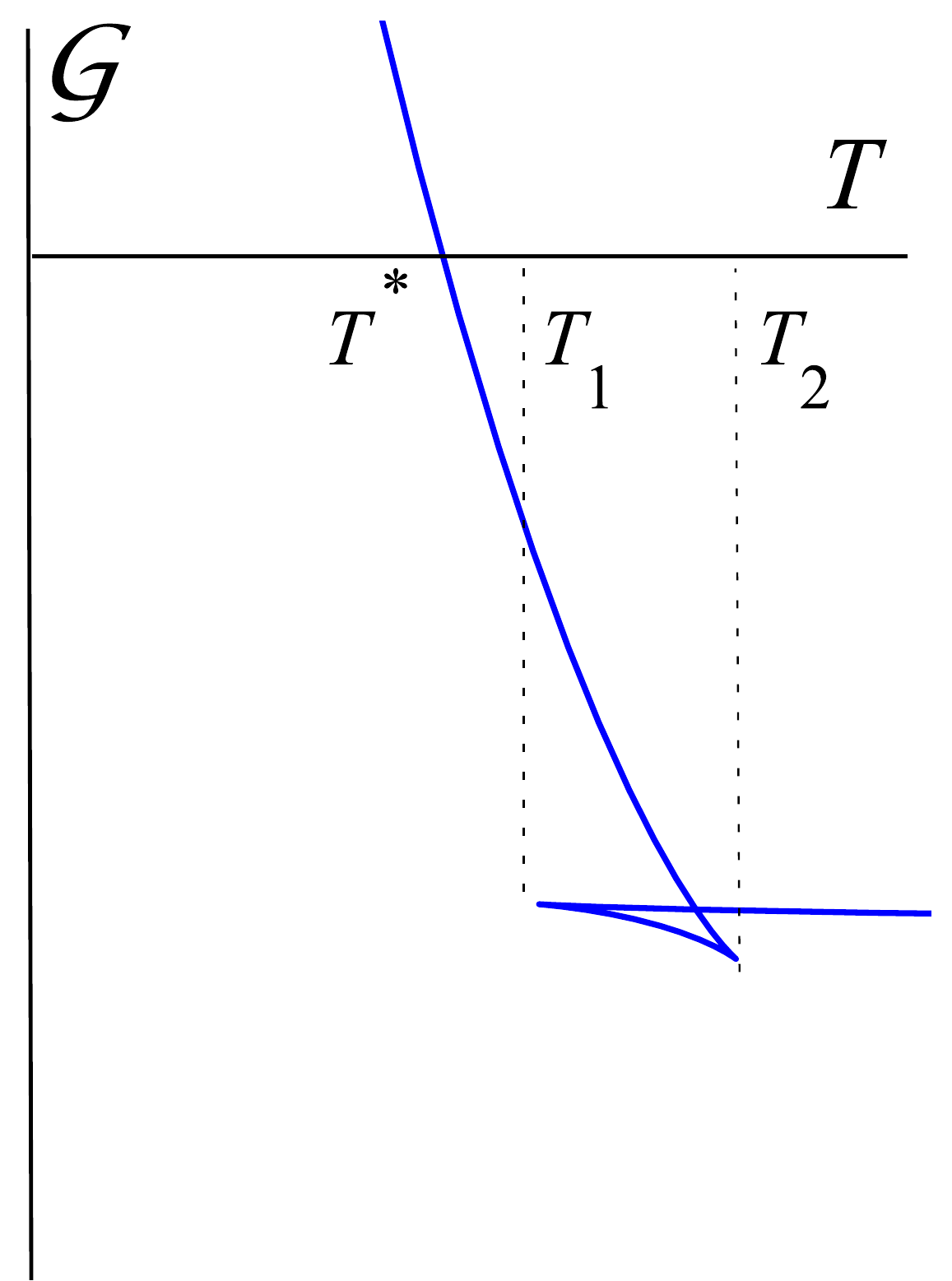}\qquad
\includegraphics[scale=0.16]{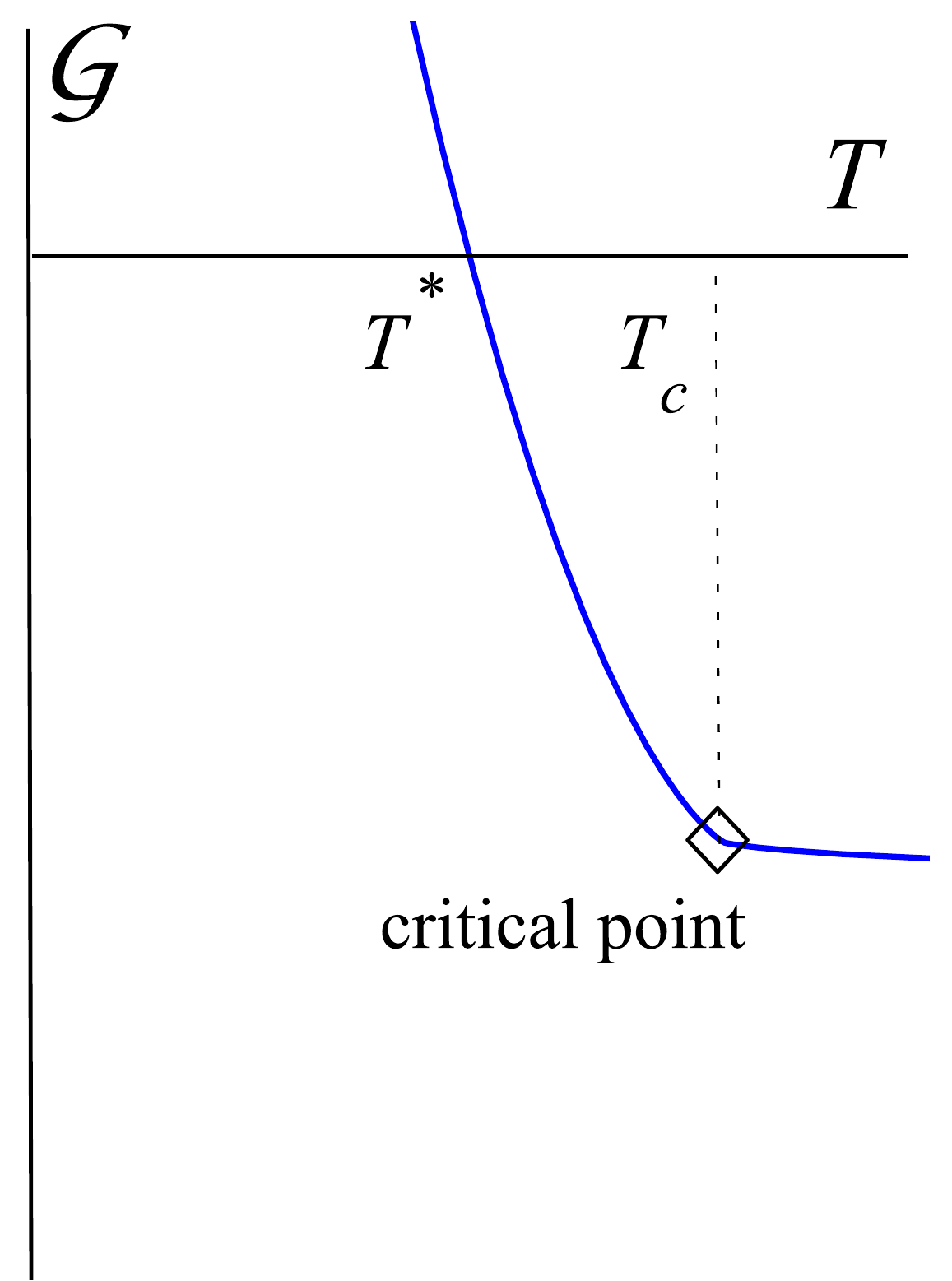}
\caption{\small $\mathcal{G}(T)$ for decreasing conjugate potentials $\Phi=\tfrac{1}{\sqrt{2}}\approx0.70711$, $0.70710$, $0.705$, and $\Phi_c\approx0.70348$. In particular, the top-right panel zooms into the physically relevant region with $T_1<T^*$, where the thermodynamically favored black hole branch dominates over flat space.} 
	\label{p4}
\end{figure}

The criticality of KK black hole with a dilaton potential is described in Figs. \ref{p2}, \ref{p3}, and \ref{p4}. The inverse swallowtail is reminiscent from the AdS case for which it is known that there exist 
reentrant phase transitions \cite{Astefanesei:2021vcp}. $T^*$ represents the temperature for which the 
black holes start to be the favorable states, though they are not thermodynamically stable. However, 
within the temperature range $T_2\geq T\geq T_1$, there exists thermodynamically stable black holes 
in flat spacetime. We would like to point out that there is a small range of the conjugate potential for which 
 $T_2\geq T^*\geq T_1$ and so this corresponds to a zero-order phase transition. 
 
Our new approach  provides a unified perspective with
which to explicitly investigate black hole criticality in various theories, e.g. when there exists a non-trivial cosmological constant and/or including higher derivative corrections \cite{Astefanesei:2020qxk, noi}.  We are going to present some related results in the near future.

\section{Acknowledgments}

This work was supported by the Fondecyt Regular Grant 1242043.

\appendix

\section{Integrability of Euclidean action}
 In this appendix, we present a general discussion of the Euclidean action's integrability. A similar computation of the Euclidean action in Einstein-Maxwell theory, without using
Einstein equations, was presented in Section 3 of  \cite{Padmanabhan:2002sha}, though we consider the general action including the counterterms that regularize it.

To obtain the thermodynamic potential and quantum statistical relation, the action should be integrable, e.g. \cite{Gibbons:1976ue, Padmanabhan:2002jr, Gibbons:2004ai} and references therein. Therefore, we could also impose a similar integrability condition for virtual spacetime geometries.

We are interested in Einstein-Maxwell-dilaton theory where the metric (\ref{ansatz}) 
is  a virtual geometry that is not necessarily a solution of Einstein equations. 
We proceed now  to compute the Euclidean action without using these 
equations of motion. Obviously, there is no a priori reason of why should be
possible to completely integrate the Euclidean action off-shell. Indeed,  when 
evaluating the Euclidean action on the virtual spacetime geometry (\ref{ansatz}), we 
have to deal with an integrability condition that depends of the theory. The integrability 
condition is a differential equation that involves only the 
conformal factor $\Omega(x)$ and  scalar field $\phi(x)$ and, once it is 
solved, it provides a different way to obtain exact hairy black hole solutions.

Let us first compute the bulk part of the action evaluating (\ref{Eaction}) on the ansatz (\ref{ansatz}). We conveniently 
arrange the final result as
\begin{equation}
\sqrt{g^\text{E}}L^\text{E}=-\frac{\sin\theta}{\eta}\[\Upsilon' 
+\eta^2\Omega\(\Omega V-2\)-2q^2e^{-\sqrt{3}\phi}
-\frac{ \Omega^2(\phi')^2+2\Omega\Omega''-3(\Omega')^2}{2\Omega}
\,f\,\]
\label{Lagr}
\end{equation}
where, to simplify its form, we introduce
\begin{equation}
	\label{ident}
	\Upsilon\equiv 2f\Omega'+\Omega f'
\end{equation}
The reason we work with this expression is that, since we have to keep $f(x)$ as an arbitrary function, we split 
the terms containing $f(x)$ in one that is integrable, $\Upsilon$, and another one proportional to 
$f(x)$ that is not. On general grounds, to be able to integrate the action, we can consider the coefficient of $f(x)$ 
in the last term of (\ref{Lagr}) as an integrability condition,
\begin{equation}\label{omeg}
	\Omega^2(\phi')^2+2\Omega\Omega''-3(\Omega')^2=0
\end{equation}

We observe that, eq. (\ref{omeg}) is in fact equivalent to the combination $\mathcal{E}^t_t-\mathcal{E}^x_x=0$, as can be explicitly verified from
\begin{equation}
\m{E}_t^t=\frac{\Omega^2(\phi')^2
+4\Omega\Omega''-3(\Omega')^2}{4\eta^2\Omega^3}f
+\frac{\Omega V-2}{2\Omega}
+\frac{2q^2e^{-\sqrt{3}\phi}+\Omega'f'}{2\eta^2\Omega^2}
\end{equation}
and
\begin{equation}
\m{E}_x^x=\frac{3(\Omega')^2-\Omega^2(\phi')^2}{4\eta^2\Omega^3}f
+\frac{\Omega V-2}{2\Omega}
+\frac{2q^2e^{-\sqrt{3}\phi}+\Omega'f'}{2\eta^2\Omega^2}
\end{equation}
yielding
\begin{equation}
\m{E}_t^t-\m{E}_x^x=\frac{\Omega^2(\phi')^2
	+2\Omega\Omega''-3(\Omega')^2}{2\eta^2\Omega^3}f
\end{equation}

Importantly, once we solve this differential equation, two of the three unknown functions, namely $\phi(x)$ and 
$\Omega(x)$, are fixed, while $f(x)$ is still an arbitrary function, which is consistent with our definition of 
virtual geometry. Due to the specific coupling between the electric field and dilaton, we choose 
the ansatz $\phi=a\ln(x)$, where $a$ is a constant that can be fixed so that $\Omega(x)$ has the simplest form,
\begin{equation}
\phi(x)=a\ln{x} \quad\Rightarrow\quad
\Omega(x)=\frac{(a^2+1)x^{\sqrt{a^2+1}-1}} {\eta^2(x^{\sqrt{a^2+1}}-1)^2}
\end{equation}
namely $a=\sqrt{3}$. 

Therefore, we get the following concrete expressions:
\begin{equation}\label{phiomega1}
	\phi(x)=\sqrt{3}\ln(x)\,, \qquad \Omega(x)=\frac{4x}{\eta^2(x^2-1)^2}
\end{equation}
We emphasize that we can obtain the exact solution by using only the integrability condition 
and one of the Einstein equations, e.g. $\mathcal{E}_t^t=0$. Replacing (\ref{phiomega1}) into $\mathcal{E}_t^t=0$,
one can integrate this equation to determine the metric function $f(x)$,
matching the exact hairy solution of \cite{Anabalon:2013qua} for $\gamma=\sqrt{3}$.

\end{document}